\documentclass{article}
\usepackage{spconf,amsmath,graphicx, float}
\usepackage{amsmath,graphicx,url,times,booktabs, tabularx}

\newlength\savedwidth
\newcommand{\wcline}[1]{\noalign{\global\savedwidth\arrayrulewidth\global\arrayrulewidth 1.0pt} \cline{#1}
\noalign{\global\arrayrulewidth\savedwidth}}

\usepackage{multirow}
\usepackage{color}

\title{Joint Acoustic and Class Inference \\ for Weakly Supervised Sound Event Detection}
\name{{\fontsize{11pt}{0pt}\selectfont Sandeep Kothinti$^{1}$, Keisuke Imoto$^{2}$, 
Debmalya Chakrabarty$^{1}$, Gregory Sell$^{3}$, Shinji Watanabe$^{1}$, Mounya Elhilali$^{1}$}\thanks{This research was supported in part by National Institutes of Health grants R01HL133043 and U01AG058532 and Office of Naval Research grants N000141612045 and N000141712736.}}
\address{$^1$ Department of Electrical and Computer Engineering, Johns Hopkins University, Baltimore, MD, USA.\\         
 $^2$ College of Information Science and Engineering, Ritsumeikan University, Shiga, Japan. \\
 $^3$ Human Language Technology Center of Excellence, Johns Hopkins University, Baltimore, MD, USA. \\}

\begin{document}

\ninept
\maketitle

\begin{sloppy}
\begin{abstract}
Sound event detection is a challenging task, especially for scenes with multiple simultaneous events. While event classification methods tend to be fairly accurate, event localization presents additional challenges, especially when large amounts of labeled data are not available. Task4 of the 2018 DCASE challenge presents an event detection task that requires accuracy in both segmentation and recognition of events while providing only weakly labeled training data. Supervised methods can produce accurate event labels but are limited in event segmentation when training data lacks event timestamps. On the other hand, unsupervised methods that model the acoustic properties of the audio can produce accurate event boundaries but are not guided by the characteristics of event classes and sound categories. We present a hybrid approach that combines an acoustic-driven event boundary detection and a supervised label inference using a deep neural network. This framework leverages benefits of both unsupervised and supervised methodologies and takes advantage of large amounts of unlabeled data, making it ideal for large-scale weakly labeled event detection. Compared to a baseline system, the proposed approach delivers a 15\% absolute improvement in F-score, demonstrating the benefits of the hybrid bottom-up, top-down approach.
\end{abstract}


\section{Introduction}
\label{sec:intro}
Everyday soundscapes present a real challenge for audio technologies that seek to parse the changing nature of the scenes and detect relevant events in the environment. With growing interest in smart devices, smart assistants and interactive technologies, there are increased efforts to develop robust ambient sound analysis systems able to detect and track different sound sources and identify events of interest. 

Parsing a scene to identify important events is a nontrivial task. Even humans exhibit a notable degree of variability in detecting occurrences of salient events when presented with realistic busy scenes \cite{Huang2017}. Machine audition has tackled the problem of sound event detection by leveraging labeled data that allow machine learning algorithms to `learn' characteristics of sound events, hence allowing the system to detect them whenever they occur \cite{virtanen2018computational}. This supervised approach yields a reasonable performance especially in constrained settings where the nature of sound events and background sounds is well captured by the labeled data available for training \cite{Cakir2017_TASLP}. In reality, however, a fully supervised approach has limited scalability especially when dealing with everyday sound environments that can vary drastically depending on the setting and density of the sources present. Acquiring large amounts of fully-labeled data in unconstrained environments is practically infeasible. The challenges of providing supervised data for event detection stem from the need to not only identify sound events in a scene, but also accurately label timestamps of occurrence of such events. 

This, in turn, raises the question of potential benefits of unlabeled data to augment supervised training methods. There is a growing number of corpora that represent various urban soundscapes, domestic or workplace environments as well as everyday sounds (e.g. \cite{Salamon2014ADA,45857}). The abundance of such labeled datasets can enrich our ability to tackle ambient sound analysis provided the right kinds of tools are available to take advantage of both labeled and unlabeled data. In its latest iteration, the DCASE 2018 task4 challenge focused on scenarios with a large amount of unlabeled data along with a small set of labeled data \cite{serizel:hal-01850270}. The availability of both data sets can be leveraged in a number of ways. A number of approaches have been proposed to supplement supervised training using unlabeled data by means of data augmentation, which can result in improved training of the machine learning systems as well as more robust event detection accuracies \cite{zhang2012semi,elizalde2017approach,han2016semi}. In parallel, unsupervised techniques have also been proposed to infer characteristics of sound events hence taking into account the dynamics of sound classes \cite{salamon2015unsupervised,Espi2015}.  

In the current work, we aim to leverage both the power of machine learning using a combination of labeled and unlabeled data to learn the characteristics of event classes, as well as our knowledge of the physical and perceptual attributes of sounds that can help guide the segmentation of sound events as they occur in a scene. The latter approach employs principles from bottom-up auditory attention models where we know changes in sound structure are flagged by the human perceptual system as salient events that attract our attention for further processing \cite{Huang2017,kim2014,Kaya2017}. Detecting the onset and offset of these events of interest provides an anchor to our event labeling system that eliminates discontinuities in event labels as well as minimizes false identification from our supervised system, resulting in notable improvement over a pure label-guided classification system. This work is an extension to our submission to DCASE 2018 \cite{Kothinti2018}, with more detailed analysis of subsystems and improved performance on the evaluation data. Section \ref{sec:related} provides an overview of related event detection systems, and focuses on prior work that has leveraged deep learning to tackle the challenge of event detection, much in the same vein as the proposed model.  Section \ref{sec:model} presents the proposed system for event detection and details the interplay between a bottom-up, acoustic-driven analysis and top-down supervised approach. The experimental setup is described in Section \ref{sec:setup} while the system performance and comparison to baseline systems is presented in Section \ref{sec:results}. Finally, Section \ref{sec:conclusion} discusses concluding remarks and future directions.

\section{Related Work}
\label{sec:related}
A large body of work related to event detection systems has focused on innovations to feature representations that provide a suitable task-relevant mapping of the raw acoustic signal (e.g. \cite{Schroder2015}). More recently, progress in deep learning methods has provided a notable jump in performance over conventional modeling methods for event detection systems (see \cite{virtanen2018computational,Mesaros2018} for a review). All best performing systems in event detection tasks in DCASE 2016 \cite{AdavannePPHV17} and DCASE 2017 \cite{DBLP:journals/corr/abs-1709-00551} include some formulation based on deep neural networks (DNN) with various flavor. While such DNN systems are generally quite powerful for their specific tasks, their performance is often limited to the exact configuration of the task itself. In \cite{DBLP:journals/corr/abs-1709-00551}, convolutional recurrent neural networks were used in a semi-supervised setting to provide notable performance improvements in segment-based evaluations, but they performed rather poorly on event-based evaluations in a similar task for DCASE 2018 \cite{serizel:hal-01850270}.

Generative models such as Restricted Boltzmann Machines (RBMs) and Conditional RBMs (cRBMs) have also been used to model audio scenes in high-dimensional representations \cite{Hinton2007-HINLML}. These models follow in the tradition of exploring robust feature representations for audio signals and can in fact reliably track multiple audio streams by encoding their regularities over new embedding spaces. In \cite{Debicassp}, mixtures of cRBMs were shown to predict unexpected events in a noisy subway station with high precision by locating time windows that deviate from an underlying statistical structure of the scene. In the present study, we employ similar models based on RBMs in order to leverage their generative nature with a focus on onset detection unlike a discriminative method \cite{8010445}. 

%
\begin{figure*}[t]
\centering
\includegraphics[width=1.7\columnwidth]{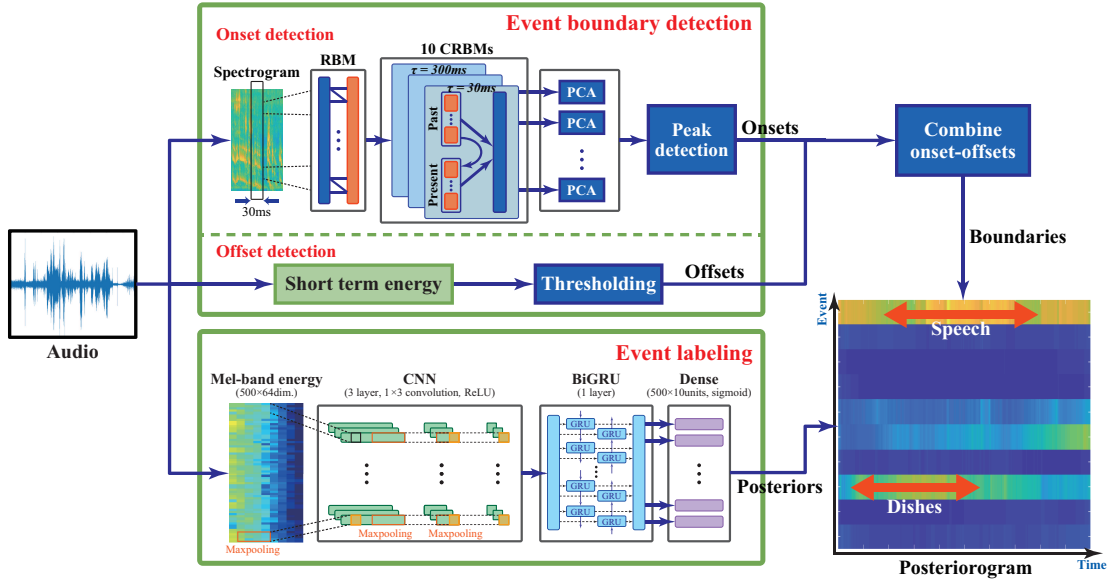}
	\caption{Proposed system for acoustic event detection. The system consists of 2 main components that operate in parallel: (i) the event boundary detection (top branch) which operates as a purely acoustic-based unsupervised analysis and yields time stamps of onsets and offsets of events of interest; and (ii) the event labeling (bottom branch) which is a supervised network trained to provide probabilistic labels of events in the input audio.}
\label{fig:CRNN}
\end{figure*}

\section{Proposed architecture}
\label{sec:model}

The proposed system combines a bottom-up (acoustic-driven) and top-down (label-guided) approach to detect sound events. Figure \ref{fig:CRNN} delineates the proposed methodology with an example. The bottom-up approach relies solely on acoustic characteristics of the audio signal to flag changes over time as captured in a high-dimensional mapping of the signal. The top-down approach is a supervised label-driven characterization of the sound labels derived from a DNN. The outputs are then combined with the bottom-up subsystem providing guidance to the windows of interest while the top-down subsystem characterizes the labels for these windows.

\subsection{Event boundary detection}
Event boundaries are identified in a purely acoustic-driven manner by tracking changes in acoustic properties of the input audio. We employ a generative framework to extract a rich mapping of the acoustic waveform that captures both local and global spectro-temporal regularities in the signal. The output of this representation is a rich array of activations in a high-dimensional space which allows tracking auditory events with different spectro-temporal characteristics. 

This acoustic analysis is structured as a hierarchical system with 3 main stages as shown the top block of figure \ref{fig:CRNN}. First, a 128 channel biomimetic auditory spectrogram $S(t,f)$ is extracted from the input audio with integration over 10ms and a frame-shift of 10ms \cite{chi2005multiresolution}.  3 consecutive frames of $S(t,f)$ are stacked to produce a temporal context of 30ms and are used as input to a Restricted Boltzmann Machine (RBM) \cite{Hinton2007-HINLML,Hinton2012a}. The RBM, trained using Contrastive Divergence (CD), is a generative model and is expected to capture local spectro-temporal dependencies of the incoming audio signal. Gaussian-Bernoulli units are used to model visible-hidden connections. After training,  RBM weights (\textbf{W}) and hidden bias (\textbf{b}) are used to transform input data (\textbf{v}) as given in (\ref{eq:rbm}).

\begin{equation}
\label{eq:rbm}
h_i=\sum_jv_jW_{ji}+b_i
\end{equation}

The next stage in the acoustic mapping further processes RBM outputs (\textbf{h}) using an array of 10 conditional RBMs \cite{taylor2007modeling,crbm}. The cRBM array further analyzes the output of the first stage along a range of temporal contexts from 30ms to 300ms, hence capturing global dynamics in the signal and tracking events with different temporal characteristics. The cRBM layer also employs Gaussian-Bernoulli visible-hidden units and is trained using CD. The weights (\textbf{W}, \textbf{A}) and biases (\textbf{b}) of the cRBM array are used as an affine transform to generate a final high-dimensional representation of the acoustic signal, as given in (\ref{eq:autoreg}). 
\begin{equation}
\begin{aligned}
\label{eq:autoreg}
b_i^t = \sum_jh_j^{t-1}A_{ji}+b_i \\
c_i^t=\sum_jh_j^tW_{ji}+b_i^t
\end{aligned}
\end{equation}

The activations across the nodes of each cRBM network are further processed using Principal Component Analysis (PCA) to get directions of maximal variance and reduce dimensionality to 16 dimensions per cRBM \cite{pca}. The PCA outputs are then processed through first-order difference and smoothed using a moving average with window length inversely proportional to the cRBM context length. The smoothed derivatives from all the dimensions are summed to produce a measure of activity in time. We flag local maxima in this activity to indicate notable changes in the acoustic signal and hence a likely index of increased acoustic event activity. The closest preceding sample at 25\% of the detected peak is marked as the onset point. In parallel, event offsets are analyzed using the short term energy (STE) of the audio signal. STE is computed using a 20ms window and is thresholded to locate low-activity points immediately following the detected onsets from the upper branch. These low-activity points paired with corresponding onsets form the event boundaries.  

\subsection{Event labeling}
\label{sec:supervised}
To label the acoustic event detected by the bottom-up approach, we employ a deep neural network trained to classify given sound classes. This neural network outputs a posterior of acoustic events for each frame, which is combined with the event boundary detection results for the class inference of acoustic events. 

\subsubsection{Convolutional recurrent neural network}
\label{sec:posteriors}
For the classification of acoustic events, we apply a convolutional recurrent neural network (CRNN), which is used as the baseline system for task 4 of DCASE 2018. This is depicted in the bottom block of fig. \ref{fig:CRNN}.
The acoustic features used in this system consist of 64-dimensional log mel-band energy extracted in 40 ms Hamming windows with 50\% overlap.
The log mel-band energy is then fed to the CRNN, which has three 2-D convolutional layers and then a layer of bi-directional gated recurrent units (BiGRU) followed by a dense layer with sigmoid activation to compute posterior probabilities of the different sounds classes. 
Pooling along the time axis is used in training with the segment-level labels, but is omitted for inference in order to yield frame-level estimations.

\vspace{-0.3cm}
\subsubsection{Label inference}
\label{sec:post_processing}
An acoustic event label is given to the unlabeled event by calculating an average posterior of each acoustic event in the active duration. The labels are chosen based on a maximum posterior criterion. An example of the labeling process is shown in fig. \ref{fig:CRNN}. 
%
\section{Experimental setup}
\label{sec:setup}
\subsection{Dataset}
\label{sec:data}
The dataset used to test this system consists of the data provided for Task4 of the DCASE Challenge 2018. It is a subset of Audioset drawn from Youtube videos and consists of various sound classes occurring in domestic contexts \cite{45857}. Training data includes 1578 audio files labeled at the segment level (referred to as weakly labeled data) along with 14,412 unlabeled in-domain 39999 out-of-domain files.  Test data consists of a development set (Dev) with 288 audio files and an evaluation set (Eval) with 880 audio files. Test data is annotated with time boundaries for each labeled event. Test files can have more than one event of the same or different class with some events even overlapping with other events.  In our system, we used only weakly labeled and unlabeled in-domain training data for both the unsupervised and supervised models. 

\subsection{Evaluation metric}
\label{sec:metric}
Event detection is evaluated event-by-event using the macro average and micro average of F-scores. Macro average is computed as the average of class-wise F-scores and micro average is the F-score of all events irrespective of classes. Error rate (ER) is used as a secondary metric to assess errors in terms of insertions, deletions, and substitutions. sed\_eval toolbox \cite{mesaros2016metrics} is used to compute F-scores and ER. Onsets are evaluated with a collar tolerance of 200ms. Tolerance for offsets is computed per event as the maximum of 200ms or 20\% of event length. An event is considered to be a hit only when the predicted label matches with the ground truth and the event boundaries correspond to the annotated boundaries. Hence any mismatch in either the labels or boundaries will result in a false positive and a false negative. 

\subsection{Baseline system}
\label{sec:baseline}
The baseline system is a CRNN with 3 CNN layers and 1 BiGRU layer, trained in two stages. During the first stage, weakly labeled data is used for training with an objective of predicting the label at clip level. 
The first trained model is used to define labels for the unlabeled in-domain data, which is then used in the second stage of training. Training progress is monitored using a held-out validation set. During the first stage of training, 20\% of weakly labeled data is used as the validation set and during the second stage of training, the entirety of the weakly labeled data is used as the validation set. 64-dimensional log Mel-band magnitudes are used as input features and the whole sound clip is given as the input to the CRNN which uses 2-D convolution in time and frequency. During test time, strong labels are assigned based on the posterior probabilities and smoothed using a median filter of length 1s.\\

\subsection{System training}
\label{sec:system}
In the proposed system, both bottom-up and top-down subsystems are trained independently. The RBM-cRBM model for event boundary detection is trained using weakly labeled and unlabeled in-domain training data. The number of hidden units is fixed at 350 for the RBM and 300 for the cRBM models. Both models are trained using a contrastive divergence-based gradient descent with 10 sampling steps. 

For the event labeling subsystem, we compare 3 variants of the top-down module described above in Section \ref{sec:posteriors}.
In System 1, event labels are predicted with a CRNN trained using only the weakly labeled data.
In System 2, the CRNN is trained using weakly labeled data (1,578 clips) and augmented data (1,080 clips) which are generated by mixing multiple weakly labeled clips.
In Systems 1 and 2, the number of channels, kernel size, stride and pooling size in the convolutional layers are \{128, 128,192\}, \{1 $\!\times\!$ 3, 1 $\!\times\!$ 3, 1 $\!\times\!$ 3\}, \{1, 1, 1\}, and \{1 $\!\times\!$ 8, 1 $\!\times\!$ 4, 1 $\!\times\!$ 2\}, and the number of GRU units is 64.
System 3 uses predictions from the DCASE 2018 baseline model for Task4. An ensemble system  uses majority vote on predictions from Systems 1-3. All parameters are tuned to maximize the performance on the development set.  
%
\section{Results}
\label{sec:results}
Table \ref{tab:result} compares the performance of the proposed systems with the baseline system for both Dev and Eval sets. As noted in the table, the proposed method improves significantly over the baseline model in terms of F-score and error rate. The Ensemble system shows the best performance and improves over the baseline by 15.18\% on Dev and 14.80\% on Eval set. Both development and evaluation sets yield similar performance improvements across all systems, validating the generalizability of the proposed method.

Looking closely at the detection scores, System 3 highlights the contributions of the acoustic-driven branch of the proposed system relative to the baseline system, since System 3 in fact utilizes the posteriors from the baseline system itself. It is also worth noting that both System 1 and System 2 are trained only using the weakly-labeled dataset. Their performance seems to indicate that labels derived from weakly labeled data are more accurate.

\begin{table}[H]
\caption{F-score and error rate in event-based metrics}
\vspace{3pt}
\label{tab:result}
\centering
\begin{tabular}{crcrc}
\wcline{1-5 \vspace{1.4pt}}
\multirow{2}{*}{{\bf Method}} & \multicolumn{2}{c}{{\bf Dev}} & \multicolumn{2}{c}{{\bf Eval}}\\
 & F-score\!&\!Error rate\!&\!F-score\!&\!Error rate\\
\wcline{1-5 \vspace{1.4pt}}
Baseline & 14.87\% & 1.52 & 10.80\% & 1.77\\
System 1 & 29.31\% & 1.40 & 23.58\% & 1.25\\
System 2 & 29.69\% & 1.44 & 23.88\% & 1.34\\
System 3 & 27.20\% & 1.46 & 23.74\% & 1.21\\
Ensemble　 & {\bf 30.05\%} & {\bf 1.36} & {\bf 25.40\%} & {\bf 1.19}\\
\wcline{1-5}
\end{tabular}
\end{table}

We look closely at the system performance across the different sound classes present in the dataset. Table \ref{tab:clwise} compares the performance for individual classes on Dev and Eval sets. To analyze the performance for different classes, we separate the events into small duration (average duration $\le$ 2s) and long duration (average duration $>$ 2s) events. This analysis sheds light on an interesting pattern. The proposed system appears to perform better compared to baseline for classes with smaller duration (marked with * in table \ref{tab:clwise}); whereas it does not yield any notable improvements relative to baseline on longer duration events. This can be attributed to two factors. Firstly, event boundaries detected using our system are more accurate for smaller duration events as these events do not have overlapping activity from other events. The acoustic-driven analysis does in fact track the statistical regularity of the incoming signal, hence allowing it to detect deviations in this regularity. The presence of overlapping events weakens the efficacy of this tracking process and subsequently the effectiveness of boundary detection on longer events. Secondly, the tolerance for error in offset is defined to be higher for longer events which minimizes the impact of boundary errors.   

\begin{table}[H]
\caption{Class-wise F-score (* marks short-duration events)}
\vspace{3pt}
\label{tab:clwise}
\centering
\begin{tabular}{crcrc}
\wcline{1-5 \vspace{1.4pt}}
\multirow{2}{*}{{\bf Class}} & \multicolumn{2}{c}{{\bf Dev}} & \multicolumn{2}{c}{{\bf Eval}}\\
 & baseline\!&\!ensemble\!&\!baseline\!&\!ensemble\\
\wcline{1-5 \vspace{1.4pt}}
Alarm/Bell* & 5.0 & 34.9 & 4.8 &43.5\\
Blender & 17.8 & 20.3 & 12.7& 24.0\\
Cat* & 0.0 & 31.2 & 2.9 & 21.9\\
Dishes* & 0.0 & 17.8 & 0.4 &12.7\\
Dog* & 0.0 & 48.1 & 2.4 & 28.9\\
Electric shaver & 35.1 & 22.6 & 20.0& 30.5\\
Frying & 29.4 & 10.5 & 24.5 & 0.0\\
Running water & 10.3 & 33.3 & 10.1 & 11.6\\
Speech* & 0.0 & 36.2 & 0.1 & 34.9\\
Vacuum cleaner & 51.1 & 45.5 & 30.2 & 45.8\\
\wcline{1-5}
\end{tabular}
\end{table}

In order to further examine the contribution of the acoustic-driven analysis to the overall event detection system, we further analyze the performance of different components of the proposed system by looking at individual modules in the pipeline. Table \ref{tab:subsystem} shows performance of onset detection, offset detection, onset-offset combination and finally overall system performance. For this analysis, we computed F-score only on the corresponding annotation. For example, for onset detection, we compute F-score excluding offset and labels and for onset-offset, we exclude labels. This comparison indicates that event labeling is poor in classifying the detected events, which deteriorates the overall performance significantly. We believe this gap can be closed by designing classification systems that are better at classifying the segments of detected events.

\begin{table}[H]
\caption{F-scores (Macro average for different subsystems)}
\vspace{3pt}
\label{tab:subsystem}
\centering
\begin{tabular}{crcrc}
\wcline{1-3 \vspace{1.5pt}}
\bf Metric & \bf Dev(F\%) & \bf Eval(F\%)\\
\wcline{1-3 \vspace{1.4pt}}
Onset only & 62.35\%  & 59.97\% \\
Offset only & 59.67\% & 54.10\% \\
Onset+Offset & 47.07\% &  41.66\% \\
Onset+Offset+Label& 30.05\% &  25.40\% \\
\wcline{1-3}
\end{tabular}
\end{table}
%
\section{Conclusion}
\label{sec:conclusion}
In this work, we propose a segmentation and recognition method for sound event detection based on joint unsupervised and semi-supervised methods. This approach combines acoustic-driven event boundary detection and supervised acoustic event classification to annotate sound events in complex acoustic scenes. One of the advantages of a parallel analysis of the incoming signal is to leverage not only known information about sound event classes (as captured by dataset labels) but also the inherent structure of these classes that distinguishes them from other classes. The use of a generative framework in the form of RBM-cRBM networks enables the tracking of these statistics in an appropriate embedding space which is subsequently used to flag deviations corresponding to new events. An interesting follow-up direction is to explore commonalities in these generative embedding spaces and those generated by the event-labeling model which is constrained by the sound classes. 

%
\bibliographystyle{IEEEtran}
\bibliography{refs}
%
%
%
%
%
%
%
%

\end{sloppy}
\end{document}